\documentclass[twocolumn,preprintnumbers,amsmath,amssymb,superscriptaddress]{revtex4}
\usepackage{amsmath,amsfonts,amssymb}
\usepackage{color}

\usepackage{amsmath,amsfonts,amssymb}
\usepackage[english]{babel}
\usepackage[T1]{fontenc}
\usepackage{color}
\usepackage{float}
\usepackage{verbatim}
\usepackage{graphicx}
\usepackage{bm}
\usepackage{mathtools}
\usepackage{stmaryrd}
\usepackage{anyfontsize}
\usepackage{changepage}

\usepackage{natbib}
\newcommand{\rr}[1]{{\rm #1}}

\newcommand{\beginsupplement}{%
        \clearpage
        \setcounter{table}{0}
        \renewcommand{\thetable}{S\arabic{table}}%
        \setcounter{figure}{0}
        \renewcommand{\thefigure}{S\arabic{figure}}%
        \setcounter{equation}{0}
        \renewcommand{\theequation}{S\arabic{equation}}
     }

\definecolor{darkblue}{rgb}{0,0,0.6}
\definecolor{darkcyan}{rgb}{0.1,0.3,0.4}
\definecolor{darkgreen}{rgb}{0,0.4,0}
\definecolor{darkred}{rgb}{0.6,0,0}

\begin{document}

\title{Diverse interactions and ecosystem engineering stabilize community assembly} 

\author{Justin D. Yeakel} \affiliation{University
  of California, Merced, Merced, CA 95340, USA} \affiliation{Santa Fe Institute, 1399 Hyde Park Road, Santa Fe, NM 87501, USA}

\author{Mathias M. Pires} \affiliation{Universidade Estadual de Campinas, Campinas - SP, Brazil}

\author{Marcus A. M. de Aguiar} \affiliation{Universidade Estadual de Campinas, Campinas - SP, Brazil}

\author{James L. O'Donnell} \affiliation{University of Washington, Seattle, WA 98195, USA}

\author{Paulo R. Guimar\~aes Jr.} \affiliation{Universidade de S\~ao Paulo, S\~ao Paulo, Brazil}

\author{Dominique Gravel} \affiliation{Universit\`e de Sherbrooke, Sherbrooke, QCJ1K0A5, Canada}

\author{Thilo Gross} \affiliation{University of California, Davis, Davis, CA 95616, USA}

\begin{abstract}
  The complexity of an ecological community can be distilled into a network, where diverse interactions connect species in a web of dependencies. Species interact not only with each other but indirectly through environmental effects, however the role of these ecosystem engineers has not yet been considered in models of ecological networks. Here we explore the dynamics of ecosystem assembly, where the colonization and extinction of species within a community depends on the constraints imposed by trophic, service, and engineering dependencies. We show that our assembly model reproduces many key features of ecological systems, such as the role of generalists during assembly, realistic maximum trophic levels, and increased nestedness with higher frequencies of mutualisms. We find that ecosystem engineering has large and nonlinear effects on extinction rates, facilitating robustness by creating niche space, but at the same time increasing the magnitude of extinction cascades. We emphasize the importance of redundancies in engineered effects and show that such redundancy lowers the barriers to colonization, promoting community diversity. Together, our results suggest that ecological engineers may enhance community diversity while increasing persistence by facilitating colonization and limiting competitive exclusion. 
\end{abstract}

\maketitle

\vspace{3mm}
\begin{adjustwidth}{2.5em}{0em}
\textbf{Significance} Community assembly is constrained by interactions between and among species, many of which can have lasting effects on the environment. We explore the influence of these ecosystem engineers on colonization and extinction dynamics using a network model that includes trophic, mutualistic, and engineering dependencies between species and the abiotic environment. We find that ecosystem engineering can stabilize assembly particularly when multiple engineers have similar effects on the community.\\
\end{adjustwidth}

To unravel nature's secrets we must simplify its abundant complexities and idiosyncrasies.
The layers of natural history giving rise to an ecological community can be distilled -- among many forms -- into a network, where nodes represent species and links represent interactions between them.
Networks are generally constructed for one type of interaction, such as food webs capturing predation \cite{Paine1966,Dunne2002,Pascual2006} or pollination networks capturing a specific mutualistic interaction \cite{Bascompte2013}, and continues to lead to significant breakthroughs in our understanding of the dynamical consequences of community structure \cite{May1972,Gross2009,Allesina2012}, assembly \cite{Ponisio2017}, and coevolution \cite{Guimaraes2017}. 
Recent interest in `multilayer networks' comprising multiple interaction types (multitype interactions) may provide additional insight into these processes \cite{Kefi2016,Pilosof2017}. 
However, interactions where species affect others by altering the environment in a lasting way have not yet been incorporated into models of ecological networks. 
These interactions, known as ecosystem engineering \cite{Lawton1994,OdlingSmee2013} or more generally niche construction \cite{OdlingSmee2013b,Fukami2015}, are quite common in nature and exist in almost every ecosystem.

Diverse interactions occur not only between species but indirectly through the effects that species have on their environment \cite{Jones1994,Olff2009,OdlingSmee2013}.
Elephants root out large saplings and small trees, enabling the formation and maintenance of grasslands \cite{Haynes2012} and creating habitat for smaller vertebrates \cite{Pringle2008}.
Burrowing rodents create shelter and promote primary production by aerating the soil \cite{Reichman2002}, salmon and aquatic invertebrates create freshwater habitats by changing stream morphology \cite{Moore2006}, and leaf-cutter ants alter microclimates, influencing seedling survival and plant growth \cite{Meyer2011}.
These examples illustrate ecosystem engineering, where the engineering organism alters the environment on timescales longer than its own \cite{Hastings2007}.


Ecosystem engineering not only impacts communities on ecological timescales, but has profoundly shaped the evolution of life on Earth \cite{Erwin2008}.
For example, the emergence of multicellular cyanobacteria fundamentally altered the atmosphere during the Great Oxidation Event of the Proterozoic roughly 2.5 Byrs BP \cite{Erwin2008,Schirrmeister2013}, paving the way for the biological invasion of terrestrial habitats.
In the oceans it is thought that rRNA and protein biogenesis of aquatic photoautotrophs drove the nitrogen:phosphorous ratio (the Redfield Ratio) to ca. 16:1 matching that of plankton \cite{Loladze2011}, illustrating that engineering clades can have much larger, sometimes global-scale effects.



%


The effect of the environment on species is commonly included in models of ecological dynamics \cite{Woodward2010,Brose2012,Gibert2019b} due to its acknowledged importance and because it can -- to first approximation -- be easily systematized. 
By comparison the way in which species engineer the environment defies easy systemization due to the multitude of mechanisms by which engineering occurs.
While interactions between species and the abiotic environment have been conceptually described \cite{Olff2009,Getz2011}, the absence of engineered effects in network models was addressed by Odling-Smee et al. \cite{OdlingSmee2013}, where they outlined a conceptual framework that included both species and abiotic compartments as nodes of a network, with links denoting both biotic and abiotic interactions.

\begin{figure}[h!]
\centering
\includegraphics[width=0.5\textwidth]{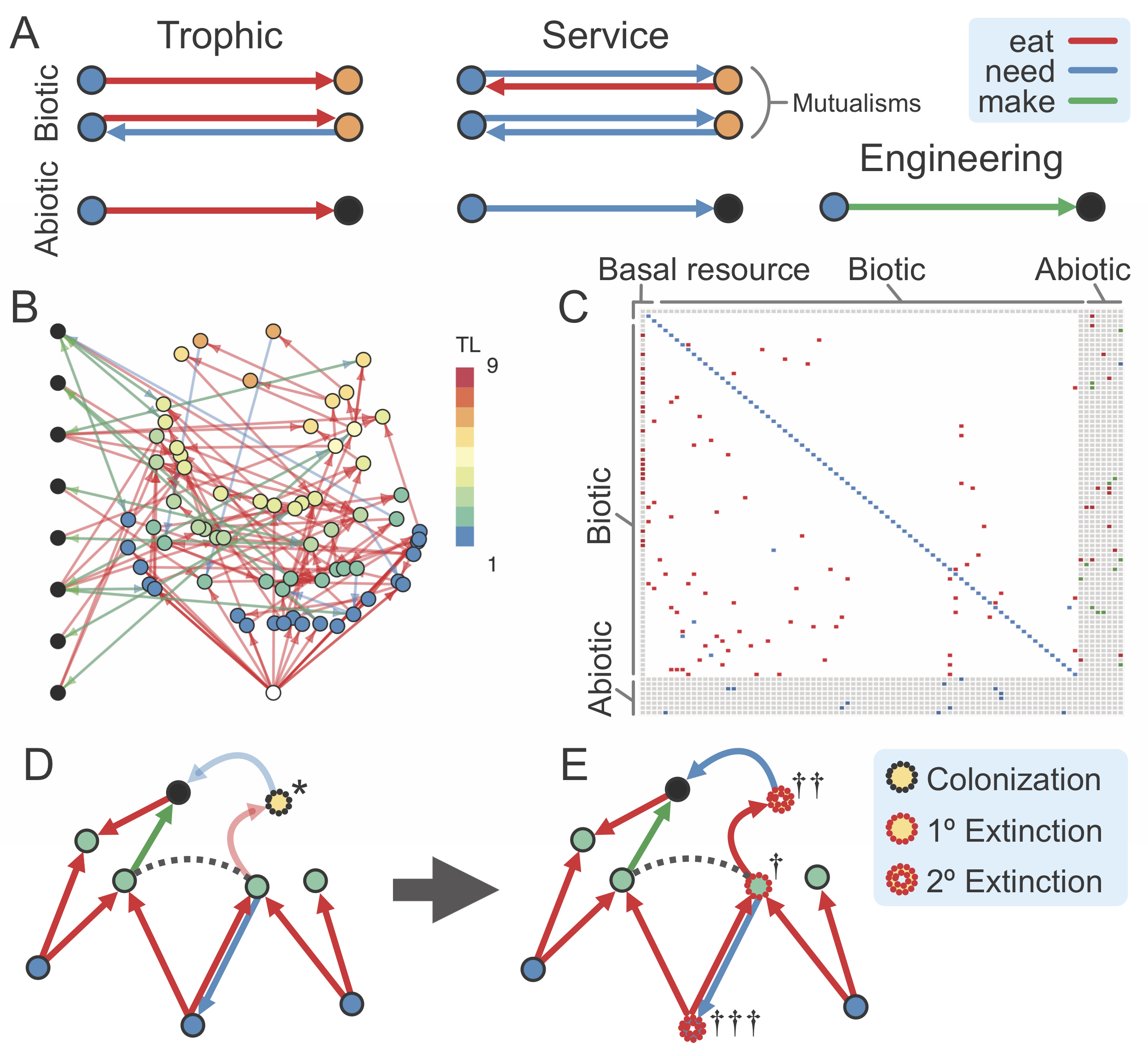}
\vspace{-6mm}
\caption{
A. Multitype interactions between species (colored nodes) and abiotic modifiers (black nodes).
B. An assembling food web with species (colored nodes; color denotes trophic level, TL) and modifiers (black nodes). The basal resource is the white node rooted at the bottom of the network.
C. The corresponding adjacency matrix with colors denoting interactions between biotic (species) and abiotic (modifiers) entities.
D. A species ($\ast$) can colonize a community when a single trophic and all service requirements are met.
E. Greater vulnerability increases the risk of primary extinction via competitive exclusion (competition denoted by dashed line) to species ($\dag$).
The extinction of species ($\dag$) will cascade to affect those connected by trophic ($\dag \dag$) and service ($\dag \dag \dag$) dependencies. 
\vspace{-3mm}
}
\vspace{-3mm}
\label{fig:model}
\end{figure}

Here we model the assembly of an ecological network where nodes represent ecological entities, including engineering species, non-engineering species, and the effects of the former on the environment, which we call abiotic \emph{modifiers}.
The links of the network that connect both species and modifiers represent trophic (\emph{eat} interactions), service (\emph{need} interactions), and engineering dependencies, respectively (Fig. \ref{fig:model}; see Materials and Methods for a full description).
Trophic interactions represent both predation as well as parasitism, whereas service interactions account for all non-trophic interactions such as pollination or seed dispersal.
In our framework a traditional mutualism (such as a plant-pollinator interaction) consists of a service (need) interaction in one direction and a trophic (eat) interaction in the other.
These multitype interactions between species and modifiers thus embed multiple dependent ecological sub-systems into a single network (Fig. \ref{fig:model}). 
Modifiers in our framework overlap conceptually with the `abiotic compartments' described in Odling-Smee et al. \cite{OdlingSmee2013}.
Following Pillai et al. \cite{Pillai2011}, we do not track the abundances of biotic or abiotic entities but only track their presence or absence.
We use this framework to explore the dynamics of ecosystem assembly, where the colonization and extinction of species within a community depends on the constraints imposed by the trophic, service, and engineering dependencies.
We then show how observed network structures emerge from the process of assembly, compare their attributes with those from empirical systems, and examine the effects of ecosystem engineers.

Our results offer four key insights into the roles of multitype interactions and ecosystem engineering in driving community assembly.
First, we show that the assembly of communities in the absence of engineering reproduces many features observed in empirical systems.
These include changes in the proportion of generalists over the course of assembly that accord with measured data and trophic diversity similar to empirical observations. 
Second, we show that increasing the frequency of mutualistic interactions leads to the assembly of ecological networks that are more nested, a common feature of diverse mutualistic systems \cite{Bascompte2003}, but are also less robust.
Our third key result shows that increasing the proportion of ecosystem engineers within a community has nonlinear effects on observed extinction rates.
While we find that a low amount of engineering increases extinction rates, a high amount of engineering has the opposite effect.
Finally we show that redundancies in engineered effects promote community diversity by lowering the barriers to colonization.\\
\vspace{-3mm}
\noindent \textbf{Assembly without ecosystem engineering}\\
\noindent Communities assemble by random colonization from a source pool.
A species from the source pool can colonize if it finds at least one resource that it can consume (one \emph{eat} interaction is satisfied; cf. Ref. \citenum{Gravel2011}) and all of its non-trophic needs are met (all \emph{need} interactions are satisfied). 
As such, the service interactions are assumed to be obligate, whereas trophic interactions are flexible.
Following the establishment of an autotrophic base, the arrival of mixotrophs and lower trophic heterotrophs create opportunities for organisms occupying higher trophic levels to invade.
This expanding niche space initially serves as an accelerator for community growth.

\begin{figure}[h!]
\centering
\includegraphics[width=0.5\textwidth]{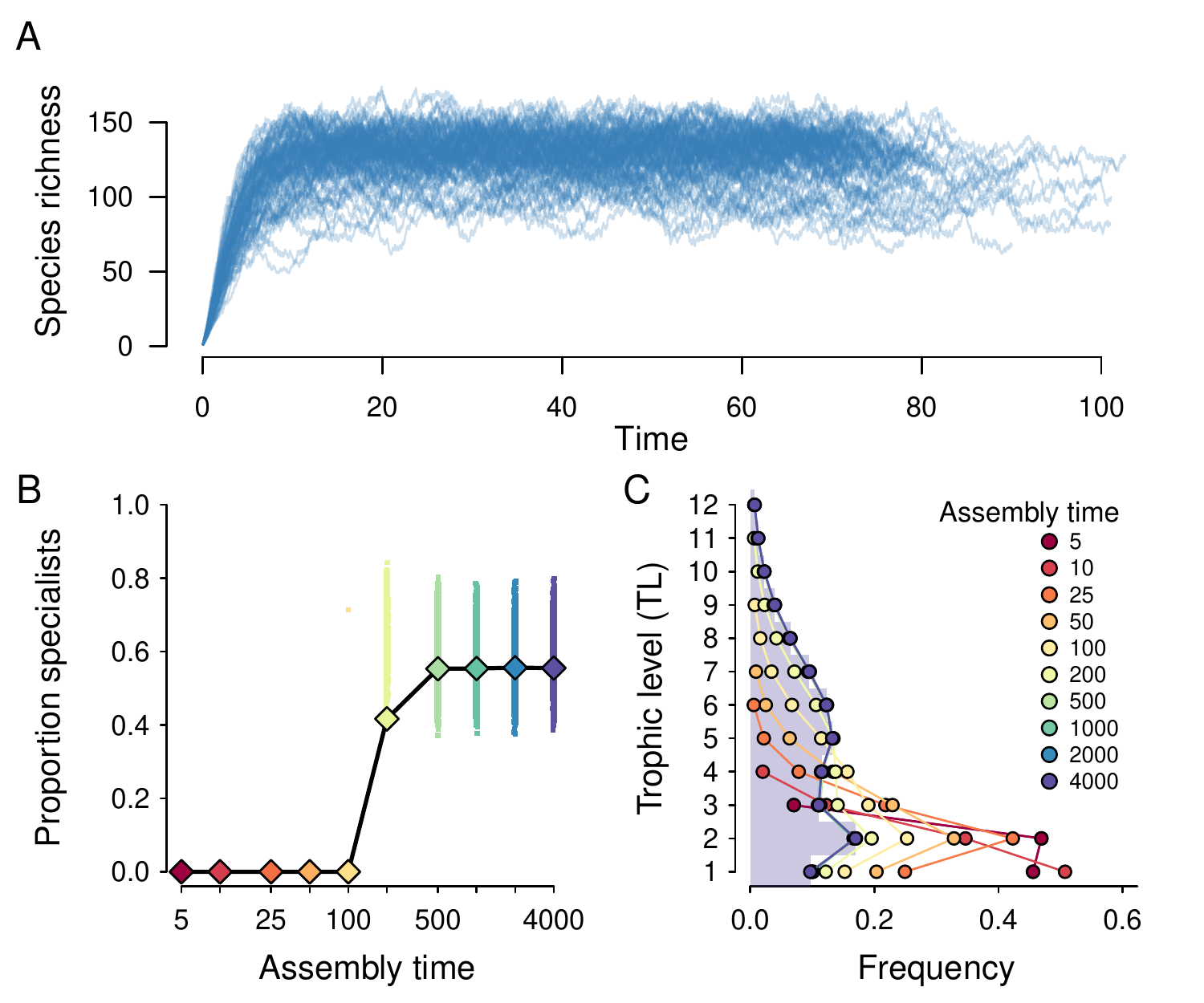}
\vspace{-6mm}
\caption{
A. Assembling communities over time from a pool of 200 non-engineering species. 
Steady state species richness is reached by $t=250$.
B. The proportion of specialists as a function of assembly time (iterations), where a specialist is defined as a species with a generality index $G_i < 1$.
All measures of $G_i$ are scaled by the average number of links per species where $L$ and $S$ are measured at steady state.
C. The frequency distribution of trophic levels as a function of assembly time (iterations). 
Autotrophs occupy ${\rm TL}=1$.
Measures were evaluated across $10^4$ replicates; see Materials and Methods for parameter values.
\vspace{-3mm}
}
\label{fig:trophic}
\end{figure}

Following the initial colonization phase, extinctions begin to slow the rate of community growth.
Primary extinctions occur by the competitive exclusion of species sharing similar resources.
A species' competition strength is determined by its interactions: competition strength is enhanced by the number of need interactions and penalized by its trophic generality (number of prey) and vulnerability (number of predators).
Secondary extinctions occur when species lose its last trophic or any of its service requirements.
See Fig. \ref{fig:model}D,~E for an illustration of the assembly process. 
As the colonization and extinction rates converge, the community reaches a steady state around which it oscillates (Fig. \ref{fig:trophic}A).
See Materials and Methods and Supporting Information, section I for a complete description of the assembly process. 

\begin{figure*}[h!]
\centering
\includegraphics[width=0.8\textwidth]{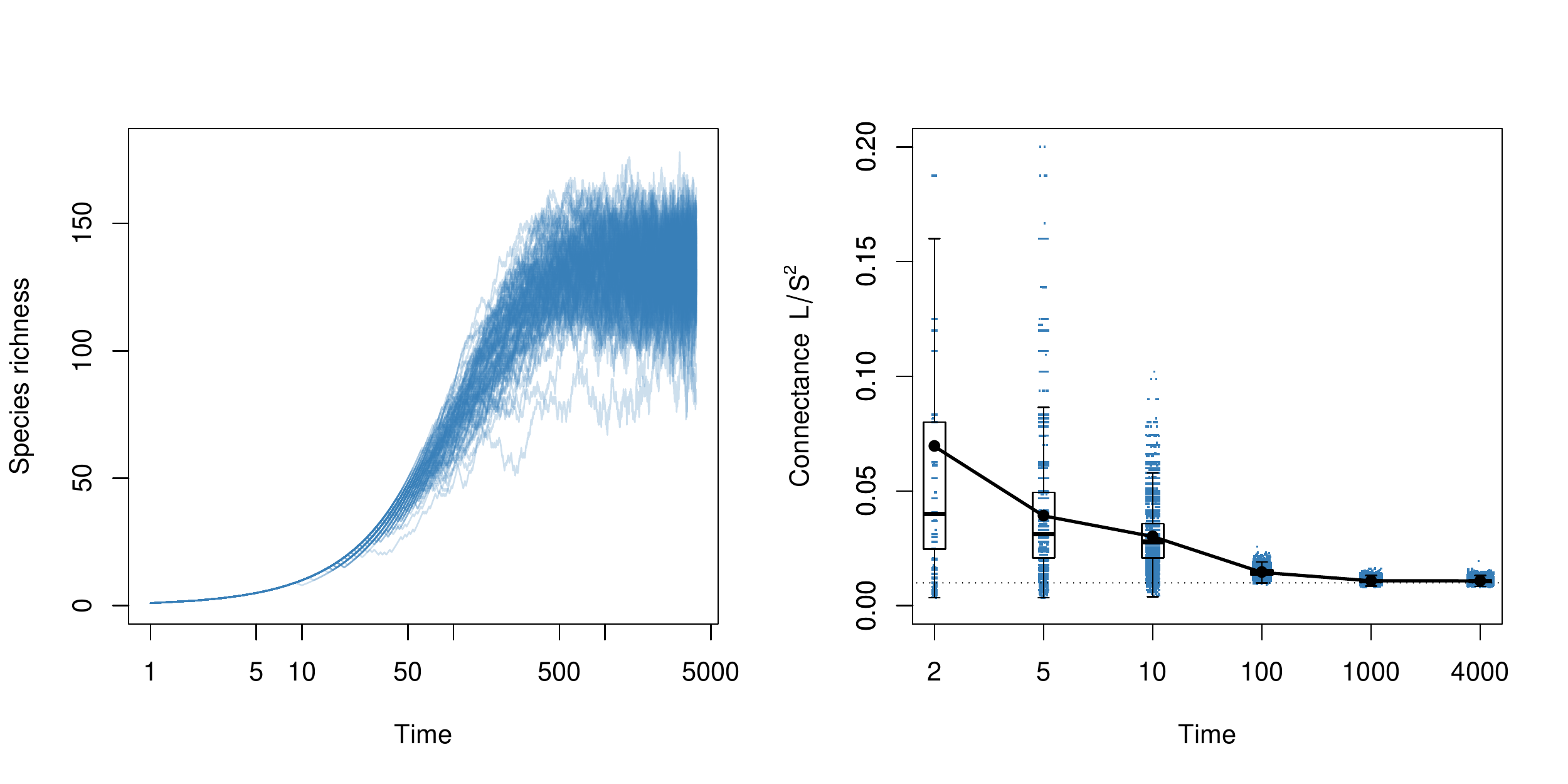}
\caption{
A. Assembly of communities over time results in steady state species richness by ca. time-step 250.
B. Trophic connectance early in assembly is high because few species are tightly connected.
Over time, connectance decays as species richness increases, and the density of trophic interactions declines.
}
\label{fig:conn}
\end{figure*}

Assembly of ecological communities in the absence of engineering results in interaction networks with structures consistent with empirical observations.
As the community reaches steady state, we find that the connectance of trophic interactions ($C=L/S^2$, where $S$ is species richness and $L$ is the number of links) decays to a value similar to that of the source pool (Fig. \ref{fig:conn}).
Decaying connectance has been documented in the assembly of mangrove communities \cite{Piechnik2008}, however this decay is a statistical inevitability, as early in the assembly process small food webs will have high link density, from which it can only decline.
Compared to trophic networks constructed using the Niche model \cite{Williams2000} given similar species richness and connectance, our framework results in networks with degree distributions of similar means but with reduced variance (Supporting Information, section II).

%

Recent empirical work has suggested that generalist species may predominate early in assembly, whereas specialists colonize after a diverse resource base has accumulated \cite{Piechnik2008,Gravel2011}.
Here the trophic generality of species $i$ is defined $G_i = k^{\rm in}_i/(L/S)$ where $k^{\rm in}_i$ is the in-degree, or number of resources consumed, by species $i$ \cite{Williams2000}.
A species is classified as a generalist if $G_i > 1$ and a specialist if $G_i < 1$.
If generalism is scaled to the steady state link density (see Supporting Information, section III), we observe that generalists dominate early in assembly, with an increase in specialists as assembly progresses (Fig. \ref{fig:trophic}B).
This confirms expectations from the trophic theory of island biogeography \cite{Gravel2011}, where early communities with lower richness are less likely to support specialist consumers than late, species-rich communities.
At steady state the proportion of specialists is ca. 56\%, similar to empirical observations of assembling food webs \cite{Piechnik2008}.

The role of specialists early in assembly is primarily due to the accumulation of autotrophic specialists.
This is evident when we observe that the trophic level (TL) distribution early in assembly ($t=5$) has an average ${\rm TL}=1.6$.
Four trophic levels are typically established by $t=50$, where colonization is still dominant, and by the time communities reach steady state the interaction networks are characterized by an average ${\rm TL_{max}}$ ($\pm$ standard deviation) $=11 \pm 2.8$ (Fig. \ref{fig:trophic}C).
While the maximum trophic level is higher than that measured in most predator-prey systems \cite{Williams2002}, it is not unreasonable if parasitic interactions (which we do not differentiate from other consumers) are included \cite{Lafferty2006}.
Overall, the most common trophic level among species at steady state is ca. ${\rm TL}=4.75$. 

The distribution of trophic levels changes shape over the course of assembly.
Early in assembly, we observe a skewed pyramidal structure, where most species feed from the base of the food web.
At steady state, we observe that intermediate trophic levels dominate, with frequencies taking on an hourglass structure (purple bars, Fig. \ref{fig:trophic}C).
Compellingly, the trophic richness pyramids that we observe at steady state follow closely the hourglass distribution observed for empirical food webs and are less top-heavy than those produced by static food web models \cite{Turney2016}.\\




\vspace{-3mm}
\noindent \textbf{Structure and dynamics of mutualisms}\\
Nested interactions, where specialist interactions are subsets of generalist interactions, are a distinguishing feature of mutualistic networks \cite{Bascompte2003}.
A nested structure has been shown to maximize the structural stability of mutualistic networks \cite{Rohr2014}, emerge naturally via adaptive foraging behaviors \cite{Valdovinos2016,Valdovinos2019} and neutral processes \cite{Krishna2008}, and promote the influence of indirect effects in driving coevolutionary dynamics \cite{Guimaraes2017}.
While models and experiments of trophic networks suggest that compartmentalization confers greater stabilizing properties \cite{Stouffer2011,Gilarranz2017}, interaction asymmetry among species may promote nestedness in both trophic \cite{Araujo2010} and mutualistic systems \cite{Pires2011}.
Processes that operate on different temporal and spatial scales may have a significant influence on these observations \cite{Massol2011}.
For example, over evolutionary time, coevolution and speciation may degrade nested structures in favor of modularity \cite{Ponisio2019}, and there is some evidence from Pleistocene food webs that geographic insularity may reinforce this process \cite{Yeakel2013}.

\vspace{-4mm}

Does the assembly of ecological networks favor nestedness when mutualistic interactions are frequent?
Increasing service dependencies (\emph{need} interactions; see Fig. \ref{fig:model}) promotes both service-resource and service-service dependencies.
Consider how species with more service interactions compare to those with fewer.
More service interactions \emph{i}) increase a species' competition strength, lowering its primary extinction risk while also \emph{ii}) increasing inter-species dependencies and its secondary extinction risk.
While mutualisms convey fitness advantages in order to evolve, the latter highlights the potential risk associated with losing mutualistic partners \cite{Bond1994,Colwell2012}.
Indeed, the balance that mutualists must maintain with their partners may have large implications for the future of global biodiversity \cite{Dunn2009}.

\begin{figure}[h!]
\centering
\includegraphics[width=0.5\textwidth]{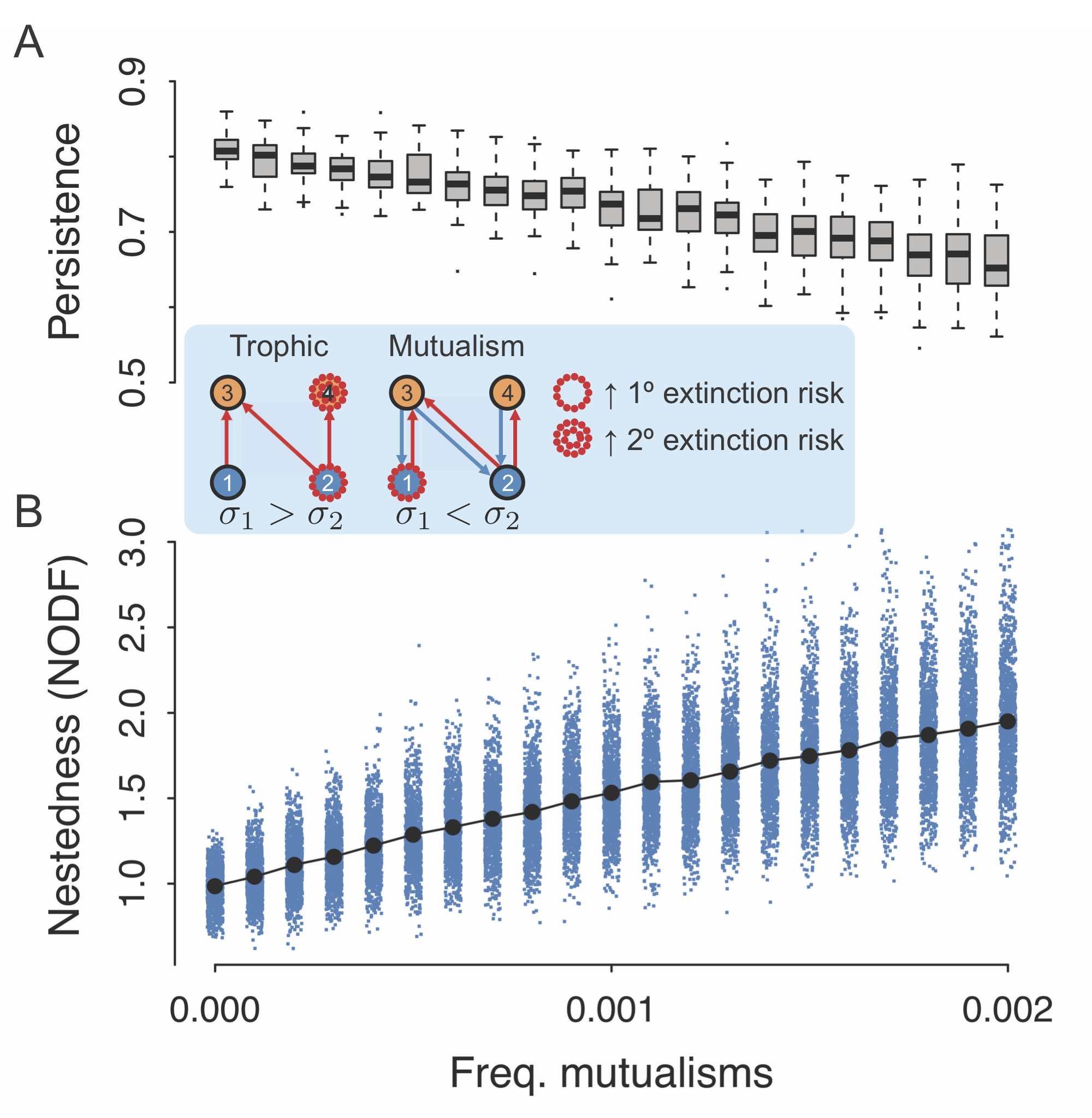}
\vspace{-8mm}
\caption{
A. Species persistence with increasing frequencies of mutualistic (service) interactions without ecosystem engineers.
B. Structural nestedness of communities, measured as NODF (Nestedness based on Overlap and Decreasing Fill) \cite{AlmeidaNeto2008}.
Measures were evaluated across $10^4$ replicates; see Materials and Methods for parameter values.
Inset: A trophic and mutualistic nested motif for resource species 1, 2 and consumer species 3, 4.
Trophic motif: the vulnerable species 2 is subject to primary extinction because it has a lower competition strength $\sigma$, resulting in an extinction cascade of species 4.
Mutualistic motif: the least vulnerable species 1 with fewer mutualistic partners is subject to primary extinction without cascading effects.
\vspace{-6mm}
}
\label{fig:nest}
\end{figure}

We find that as we increase the frequency of mutualistic interactions, the assembled community at steady state becomes more nested (Fig. \ref{fig:nest}).
In this case, nestedness emerges from the assembly process and provides structural robustness.
The robustness can be observed by examining the exclusionary differences between species in a simple nested motif (Fig. \ref{fig:nest}, inset).
In the trophic motif, species with high vulnerability (multiple predators) are at greater risk of primary extinction via competitive exclusion.
This will result in the secondary extinction of the specialist consumer, rendering the nested structure prone to change.
In our framework mutualistic networks are generally formed by composite interactions, where the consumer species is engaged in a trophic interaction while the resource species is engaged in a service interaction.
As such, the consumer species becomes a trophic partner and the resource species gains the competitive advantages of the service.
If the competitive advantages of services are greater than the costs of vulnerability (see Materials and Methods), it is the low vulnerability species with fewer trophic partners that is at greater risk of exclusion (Fig. \ref{fig:nest}, inset).
Because its elimination will not cascade, the nested structure will be more resistant to change.

Our results also suggest that the addition of mutualistic interactions comes at a cost to the assembling community.
Because mutualisms increase dependencies between species, they also increase the frequency of secondary extinctions (Fig. \ref{fig:nest}).
Measuring persistence in terms of the proportion of time species are established in the network reveals that more frequent mutualisms leads on average to lower persistence.
At the community-scale, lower average persistence implies greater species turnover.
Observations of empirical systems appear to support our model predictions.
For example, assembling plant-pollinator systems have demonstrated high rates of species and interaction turnover, both during the assembly process and at the steady state \cite{Ponisio2017}.

We emphasize that we have restricted ourselves to examining the effects of obligate mutualisms, although the importance of non-obligate mutualisms has long been recognized \cite{Ramos2012,Vieira2015,Valdovinos2016,Ponisio2017,Valdovinos2019}.
While the inclusion of non-obligate mutualists will lower the likelihood of cascading effects in systems with higher frequencies of service interactions, the loss of obligate mutualistic partners will have larger dynamic consequences than the loss of more flexible non-obligate mutualistic partners.
As such, we do not expect inclusion of non-obligate mutualisms to alter the qualitative nature of our findings.\\

\vspace{-3mm}
\noindent \textbf{Assembly with ecosystem engineering}\\
The concept of ecosystem engineering, or more generally niche construction, has both encouraged an extended evolutionary synthesis \cite{Laland2015} while also garnering considerable controversy \cite{Gupta2017,Feldman2017}.
Models that explore the effects of ecosystem engineering are relatively few, but have covered important ground \cite{Hastings2007,OdlingSmee2013}.
For example, engineering has been shown to promote invasion \cite{Cuddington2004}, alter primary productivity \cite{Wright2004}, and change the selective environment over eco-evolutionary timescales \cite{Kylafis2008,Krakauer2009} which can lead to unexpected outcomes such as the fixation of deleterious alleles \cite{Laland1999}.
On smaller scales, microbiota construct shared metabolitic resources that have a significant influence on microbial communities \cite{Kallus2017}, the dynamics of which may even serve as the missing ingredient stabilizing some complex ecological systems \cite{Muscarella2017}.

We next explore the effects of ecosystem engineering by allowing species to produce abiotic modifiers as additional nodes in the ecological network (Fig. \ref{fig:model}).
These modifier nodes produced by engineers can serve to fulfill resource or service requirements for other species.
The parameter $\eta$ defines the mean number of modifiers produced per species, drawn from a Poisson distribution (see Materials and Methods for details).
Increasing the frequency of engineering interactions both increases the number of engineering species (those species making $\geq 1$ modifier) and the number of modifiers per species.
There are two characteristics of engineering that have particular relevance for community assembly:
modifiers can linger in the community even after the species that produce them have been excluded, and
more than one engineer can produce the same modifier such that engineering redundancies increase with $\eta$ (Supporting Information, section I).



Increasing engineering has significant consequences for community robustness, but these effects also are sensitive to the frequency of service interactions within the community.
We measure community robustness by 
\emph{i}) rates of primary versus secondary extinctions,
\emph{ii}) species persistence, and 
\emph{iii}) steady state community diversity.
All measures were averaged over each species within the community across assembly time.

\begin{figure}[h!]
\centering
\includegraphics[width=0.5\textwidth]{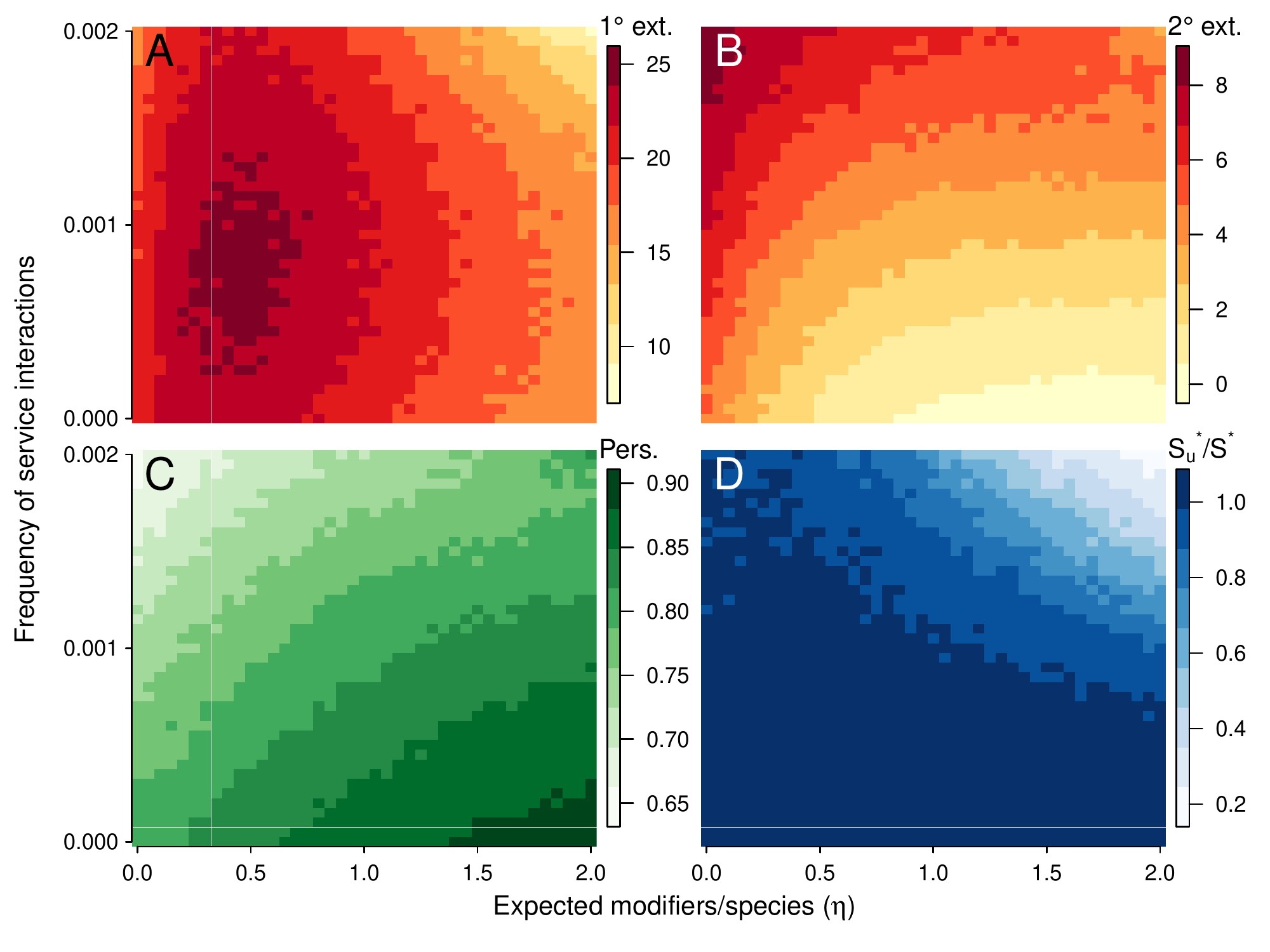}
\vspace{-8mm}
\caption{
Community robustness as a function of the frequency of service interactions and modifiers per species.
A. Mean rates of primary extinction, where primary extinctions occur from competitive exclusion of consumers over shared resources.
B. Mean rates of secondary extinction, which cascade from primary extinctions.
C. Mean species persistence.
D. The ratio $S^*_{\rm u}/S^*$, where $S^*_{\rm u}$ denotes steady states for systems where all engineered modifiers are unique to each engineer, and $S^*$ denote steady states for systems with redundant engineering. Lower values of $S^*_{\rm u}/S^*$ mean that systems with redundant engineers have higher richness at the steady state than those without redundancies.
Measures were evaluated across $50$ replicates; see Materials and Methods for parameter values.
\vspace{-5mm}
}
\label{fig:engineers}
\end{figure}

As the number of engineers increase, mean rates of primary extinction are first elevated and then decline (Fig. \ref{fig:engineers}A).
This nonlinear effect of engineering on rates of primary extinction results from two competing forces.
Increased production of abiotic modifiers supplies consumers with additional resources, limiting secondary extinctions and promoting persistence (Fig. \ref{fig:engineers}B,~C).
However, the stabilization of consumers ultimately results in increased vulnerability of prey.
Engineering dependencies are considered rare when the ratio of modifier nodes per species is $0 < \eta \leq 0.5$.
The cumulative effect in these species-rich/modifier-poor systems is increased competitive exclusion of prey and higher rates of primary extinction (Fig. \ref{fig:engineers}A).
Notably the presence of even a small number of engineers serves to limit the magnitude of secondary extinction cascades.
Higher rates of primary extinction coupled with lower rates of secondary extinction mean that extinctions are common, but of limited magnitude such that disturbances are compartmentalized.
As engineering becomes common ($\eta > 0.5$) the available niche space expands, lowering competitive overlap and suppressing both primary and secondary extinctions.

Increasing the frequency of service interactions promotes service interactions between species and engineered modifiers (Fig. \ref{fig:model}).
A topical example of the latter is the habitat provided to invertebrates by the recently discovered rock-boring teredinid shipworm (\emph{Lithoredo abatanica}) \cite{Shipway2019}.
Here, freshwater invertebrates are serviced by the habitat modifications engineered by the shipworm, linking species indirectly via an abiotic effect (in our framework via a modifier node).
As the frequency of service interactions increases, the negative effects associated with rare engineers is diminished (Fig. \ref{fig:engineers}A).
Increasing service interactions both elevates the competitive strength of species receiving services (from species and/or modifiers), while creating more interdependencies between and among species.
As trophic interactions are replaced by service interactions, previously vulnerable species gain a competitive foothold and persist (Fig. \ref{fig:nest}, inset), lowering rates of primary extinctions (Fig. \ref{fig:engineers}A). 
The costs of these added services to the community are an increased rate of secondary extinctions (Fig. \ref{fig:engineers}B) and higher species turnover (Fig. \ref{fig:engineers}C).
Low rates of primary extinction coupled with high rates of secondary extinction mean that extinctions are less common but lead to larger cascades.



\vspace{-4mm}

While the importance of engineering timescales has been emphasized previously \cite{Hastings2007}, redundant engineering has been assumed to be unimportant \cite{Lawton1994}.
We argue that redundancy may be an important component of highly engineered systems, and particularly relevant when there exists a positive feedback between the effects of engineers on their fitness \cite{Cuddington2004}.
The vast majority of contemporary ecosystem engineering case studies focus on single taxa, such that redundant engineers appear rare \cite{Lawton1994}.
However if we consider longer timescales, increasing diversity of engineering clades may promote redundancy, and in some cases this may feed back to accelerate diversification \cite{OdlingSmee2013b}.
Such positive feedback mechanisms likely facilitated the global changes induced by cyanobacteria in the Proterozoic \cite{Erwin2008,Schirrmeister2013} among other large-scale engineering events in the history of life \cite{Erwin2008}.
Engineering redundancies are likely important on shorter timescales as well.
For example, diverse sessile epifauna on shelled gravels in shallow marine environments are facilitated by the engineering of their ancestors, such that the engineered effects of the clade determine the future fitness of descendants \cite{Kidwell1986}.
In the microbiome, redundant engineering may be very common due to the influence of horizontal gene transfer in structuring metabolite production \cite{Polz2013}.
In these systems, redundancy in the production of shared metabolitic resources may play a key role in community structure and dynamics \cite{Kallus2017,Muscarella2017}.

\begin{figure*}[h!]
\centering
\includegraphics[width=0.8\textwidth]{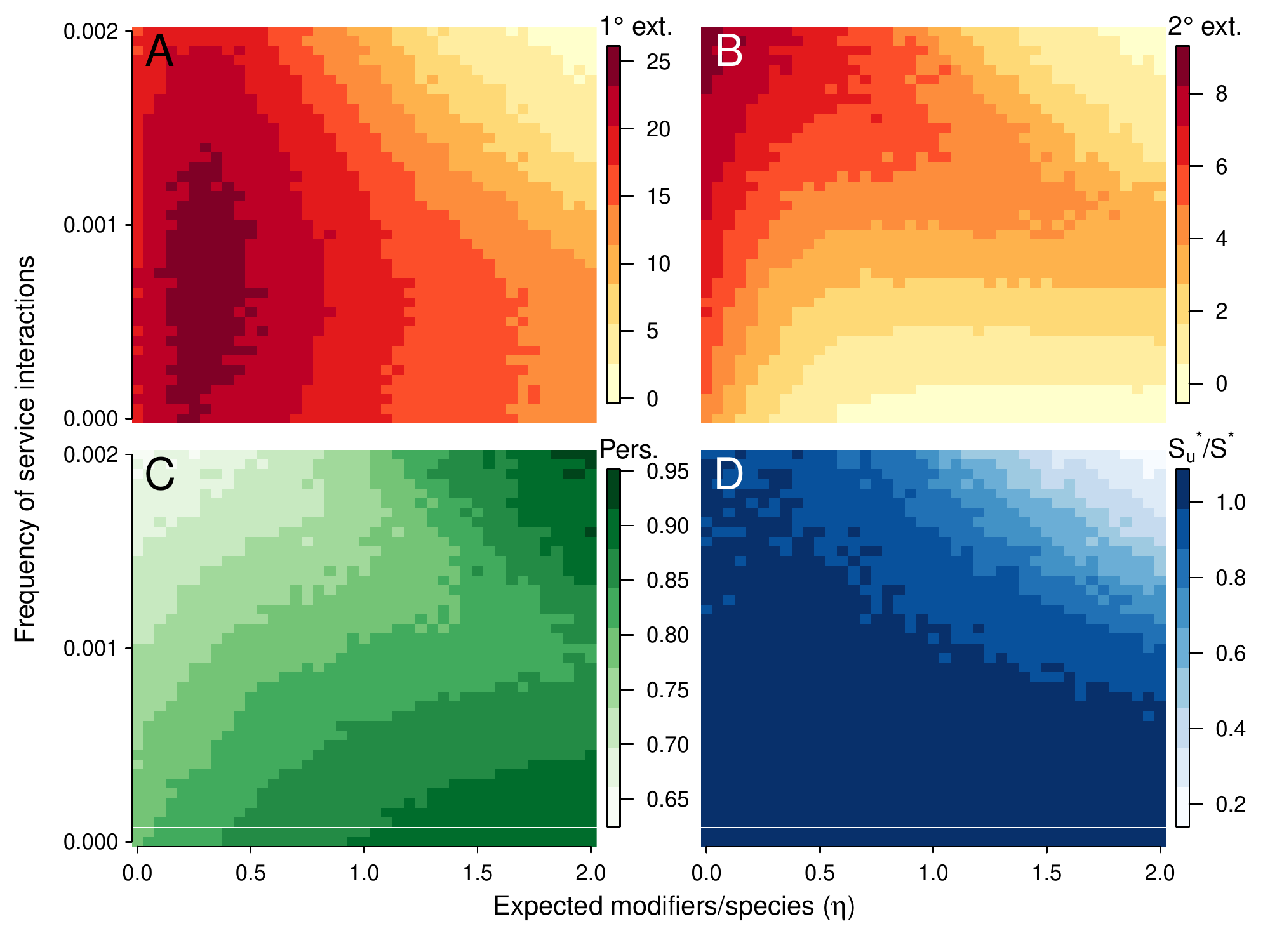}
\caption{
Measures of community robustness as a function of the frequency of service interactions and number of modifiers per species, where each modifier is uniquely made by an engineer.
A. Mean rates of primary extinction, where primary extinctions occur from competitive exclusion of consumers over shared resources.
B. Mean rates of secondary extinction, which cascade from primary extinctions.
C. Mean species persistence, defined as the percent simulation time the community is occupied by a given species, averaged across all species that successfully colonize.
D. The ratio $S^*_{\rm u}/S^*$, where $S^*_{\rm u}$ denotes steady states for systems where all engineered modifiers are unique to each engineer, and $S^*$ denote steady states for systems with redundant engineering. Lower values of $S^*_{\rm u}/S^*$ mean that systems with redundant engineers have higher steady states than those without redundancies.
Values are averaged over 50 replicates for each parameterization.
See Materials and Methods for default parameter values.
}
\label{fig:unique}
\end{figure*}

\begin{figure*}[h!]
\centering
\includegraphics[width=0.8\textwidth]{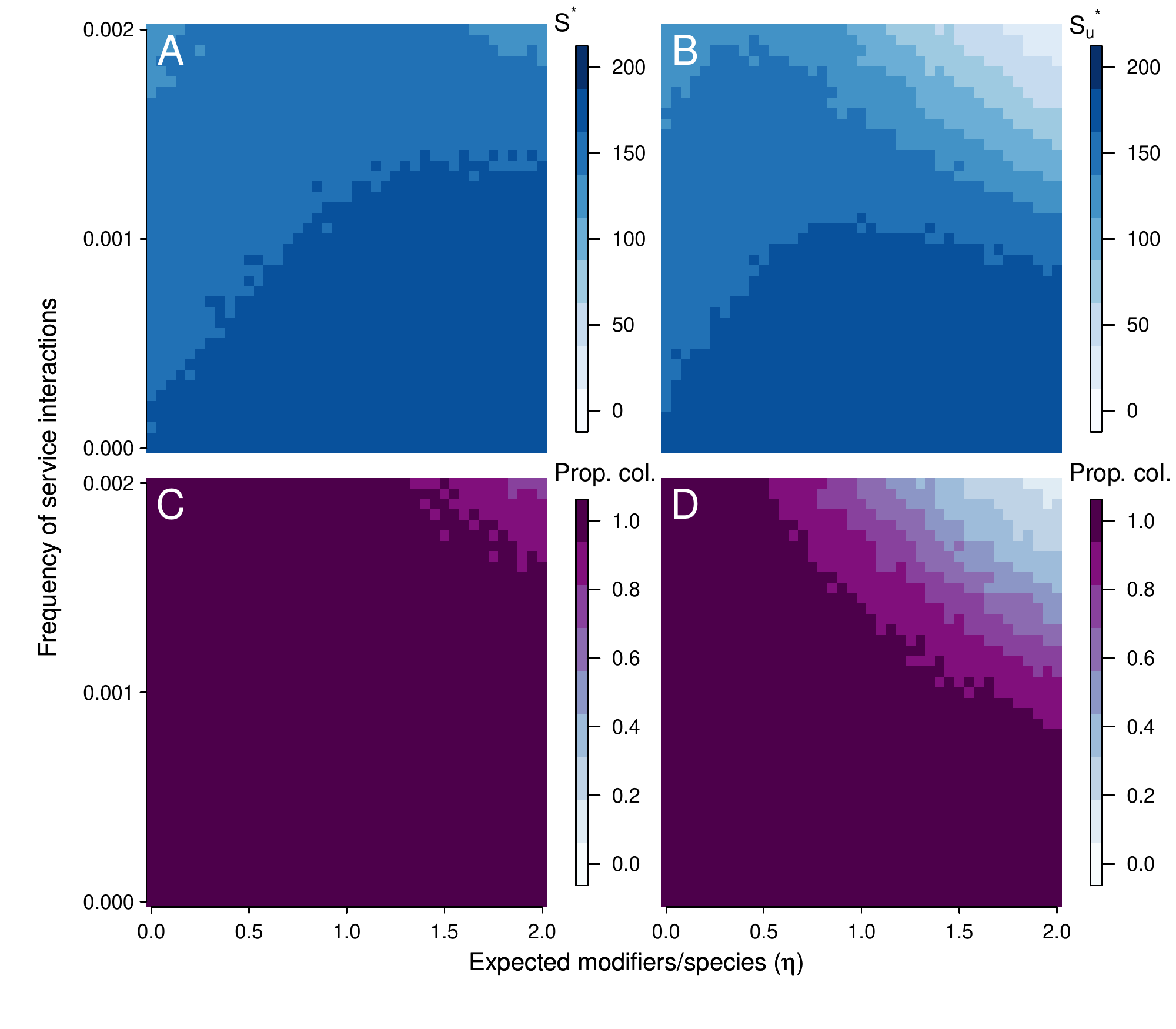}
\caption{
A. Steady state community richness with redundant engineering.
B. Steady state community richness without redundant engineering.
C. Proportion of species in the source pool that colonize the community at least once throughout the simulation (with redundant engineering).
D. Proportion of species in the source pool that colonize the community at least once throughout the simulation (without redundant engineering).
}
\label{fig:steadystate}
\end{figure*}

When there are few engineers, each modifier in the community tends to be unique to a particular engineering species.
Engineering redundancies increase linearly with $\eta$ (Supporting Information, section I; Fig. \ref{fig:redundancy}), such that the loss of an engineer will not necessarily lead to the loss of engineered modifiers. 
We examine the effects of this redundancy by comparing our results to those produced by the same model, but where each modifier is uniquely produced by a single species.
Surprisingly, the lack of engineering redundancies does not alter the general relationship between engineering and measures of community robustness (Fig. \ref{fig:unique}).
However we find that redundancies play a central role in maintaining species diversity.
When engineering redundancies are allowed, steady state community richness $S^*$ does not vary considerably with increasing service interactions and engineering (Fig. \ref{fig:steadystate}A).
In contrast, when redundant engineering is not allowed, steady state community richness $S^*_u$ declines sharply (Figs. \ref{fig:engineers}D, \ref{fig:steadystate}B).

Communities lacking redundancy have lower species richness because sparse interdependencies preclude colonization (Fig. \ref{fig:steadystate}C,D).
Colonization occurs only when trophic and service dependencies are fulfilled.
A species requiring multiple engineered modifiers, each uniquely produced, means that each required entity must precede colonization.
This magnifies the role of priority effects in constraining assembly order \cite{Fukami2015}, precluding many species from colonizing.
In contrast, redundancy increases the niche space available to species while minimizing priority effects by allowing multiple engineers to fulfill dependencies.
Our results thus suggest that redundant engineers may play important roles in assembling ecosystems by lowering the barriers to colonization thereby promoting community diversity.

Together, the results of our model point to the importance of considering multitype interactions both between species and as mediated through changes to the environment via engineering.
We suggest that including the effects of engineers, either explicitly as we have done here, or otherwise, is vital for understanding the inter-dependencies that define ecological systems.
As past ecosystems have fundamentally altered the landscape on which contemporary communities interact, future ecosystems will be defined by the influence of engineering today.
Understanding the role of ecosystem engineers is thus tantamount to understanding our own.\\



\vspace{-2mm}
\noindent \textbf{Materials and Methods}\\
  \footnotesize{
  We model an ecological system with a network where nodes represent \emph{ecological entities} such as populations of species and or the presence of abiotic modifiers affecting species such as (examples).
  Following Pilai et al. \cite{Pillai2011}, we do not track the abundances of entities but only track their presence or absence.
  The links of the network represent interactions between pairs of entities (x,y).
  We distinguish three types of such interactions: x eats y, x needs y to be present, x makes modifier y.

  The assembly process entails two steps: first a source pool of species is created, followed by colonization/extinction into/from a local community.
  The model is initialized by creating $S$ species and $M = \eta S$ modifiers, such that $N=S+M$ is the average total number of entities and $\eta$ is the average number of modifiers per species in the system.
  For each pair of species (x,y) there is a probability $p_e$ that x eats y and probability $p_n$ that x needs y.
  For each pair of species x and modifier m, there is a probability $q_e$ that x eats m and a probability $q_n$ that species x needs modifier m.
  Additionally, each species makes a number of modifiers that is drawn from a Poisson distribution with mean $\mu = \eta e/(e-1)$ where $e$ is Euler's number.
  Once the number of modifiers per species is determined, each modifier is assigned to a species independently.
  This means that multiple species may make the same modifier, and that there may be some modifiers that are not made by any species, which are eliminated from the pool.

  In addition to interactions with ecosystem entities, there can be interactions with a basal resource, which is always present.
  The first species always eats this resource, such that there is always a primary producer in the pool.
  Other species eat the basal resource with probability $p_e$.
  Species with zero assigned trophic interactions are assumed to be primary producers.
  See Supporting Information, section I for additional details on defining the source pool.

  We then consider the assembly of a community which at any time will contain a subset of entities in the pool and always the basal resource.
  In time, the entities in the community are updated following a set of rules.
  A species from the pool can colonize the community if the following conditions are met:
  1) all entities that a species needs are present in the community, and
  2) at least one entity that a species eats is present in the community.
  If a colonization event is possible, it occurs stochastically in time with rate $r_\rr{c}$.

  An established species is at risk of extinction if it is not the strongest competitor at least one of its resources that it eats.
  We compute the competitive strength of species $i$ as
  \begin{equation}
    \sigma_i = c_\rr{n} n_i - c_\rr{e} e_i - c_\rr{v} v_i,
  \end{equation}
  where $n_i$ is the number of entities that species $i$ needs, $e_i$ is the number of entities from the pool that species $i$ can eat, and $v_i$ is the number of species in the community that eat species $i$.
  This captures the ecological intuition that mutualisms provide a fitness benefit, specialists are stronger competitors than generalists, and many predators entail an energetic cost.
  The coefficients $c_\rr{n},~c_\rr{e},~c_\rr{v}$ describe the relative effects of these contributions to competitive strength.
  In the following, we use the values $c_\rr{n} > c_\rr{e} > c_\rr{v}$, such that the competitive benefit of adding an additional mutualism is greater than the detriment incurred by adding another prey or predator.
  A species at risk of extinction leaves the community stochastically in time at rate $r_e$.

  A modifier is present in the community whenever at least one species that makes the modifier is present.
  If a species that makes a modifier colonizes a community, the modifier is created immediately, however modifiers may persist for some time after the last species that makes the modifier goes extinct.
  Any modifier that has lost all of its makers disappears stochastically in time at rate $r_m$.

  The model described here can be simulated efficiently with an event-driven simulation utilizing a Gillespie algorithm.
  In these types of simulations, one computes the rates $r_j$ of all possible events $j$ in a given step.
  One then selects the time at which the next event happens by drawing a random number from an exponential distribution with mean $1/\sum_j{r_j}$.
  At this time, an event occurs that is randomly selected from the set of possible events such that the probability of event $a$ is $r_a/\sum_j{r_j}$.
  The effect of the event is then realized and the list of possible events is updated for the next step.
  This algorithm is known to offer a much better approximation to the true stochastic continuous time process than a simulation in discrete time steps, while providing a much higher numerical efficiency \cite{Gillespie1977}.
  Simulations described in the main text have default parameterizations of $S=200$, $p_{\rm e}=0.01$, $c_{\rm n} = \pi$, $c_{\rm e} = \sqrt{2}$, $c_{\rm v} = 1$, and $4000$ iterations.}

\vspace{0mm}
\noindent \textbf{Acknowledgements}\\
  \footnotesize{
  We would like to thank
  Uttam Bhat,
  Irina Birskis Barros,
  Emmet Brickowski,
  Jean Philippe Gibert,
  Chris P Kempes,
  Taran Rallings,
  Samuel Scarpino,
  Megha Suswaram,
  and Ritwika VPS 
  for insightful discussions and comments throughout the lengthy gestation of this manuscript.
  The original idea was conceived at the Networks on Networks Working Group in G\"ottingen, Germany (2014) and the Santa Fe Institute (2015).
  This work was formerly prepared as a part of the Ecological Network Dynamics Working Group at the National Institute for Mathematical and Biological Synthesis (2015-2019), sponsored by the National Science Foundation through NSF Award DBI-1300426, with additional support from The University of Tennessee, Knoxville.
  Infinite revisions were conducted at the Santa Fe Institute made possible by travel awards to JDY and TG.
  Additional support came from UC Merced startup funds to JDY, the International Centre for Theoretical Physics ICTP-SAIFR, FAPESP (2016/01343-7) and CNPq (302049/2015-0) to MAMA, CNPq and FAPESP (2018/14809-0) to PRG, and DFG research unit 1748 and EPSRC (EP/N034384/1) to TG.
  }
\clearpage

\clearpage

\beginsupplement

\section*{Supplementary Information}

%

\subsection*{Section I: Building the source pool}
Here and henceforth, we refer to the assembly model presented in the main text as the ENIgMa model (E:eat, N:need, Ig:ignore, Ma:make).
To initiate the ENIgMa assembly model, we must first constract the source pool, where each ecological entity (species and modifiers) is defined by its potential interactions with each other.
The source pool interaction matrix $\bm P$ is generated by first setting the number of species in the pool $\mathcal{S}_{\bm P}$ and determining the number of modifiers $\mathcal{M}_{\bm P}$ that are made by ecosystem engineers.
The resulting matrix is $\mathcal{N}_{\bm P}\times\mathcal{N}_{\bm P}$ where $\mathcal{N}_{\bm P}=\mathcal{S}_{\bm P}+\mathcal{M}_{\bm P}$, and is subdivided into four quadrants, only two of which play a role here: species-species interactions and species-modifier interactions (see Fig. \ref{fig:model}).
In these two quadrants, the expected frequency of eat interactions ${\rm E}\{p_\rr{e}\}$ and the expected frequency of need interactions ${\rm E}\{p_\rr{n}\}$ are free parameters, as is the expected number of modifiers made per species ${\rm E}\{\mathcal{M}_i\}=\eta$.
Here and throughout, we simplify this parameter space by assuming that the frequency of eat and need interactions for species-species (SS) interactions and species-modifier (SM) interactions are equivalent, such that ${\rm E_{SS}}\{p_\rr{e}\} = {\rm E_{SM}}\{p_\rr{e}\}$ and ${\rm E_{SS}}\{p_\rr{n}\} = {\rm E_{SM}}\{p_\rr{n}\}$.
For each species, a set number of modifiers is drawn from ${\rm Poiss}(\eta)$, such that the expected proportion of species that are engineers (species that make modifiers) is $1-{e}^{-\eta}$.
If a particular modifier is randomly and independently drawn for a given engineer from a complete list of all possible modifiers, such that multiple species -- with some probability -- can make the same modifier, the expected number of modifiers is
\begin{equation}
{\rm E}\{\mathcal{M}_{\bm P}\} = \mathcal{S}_{\bm P}\eta\left(1 - \frac{1}{{e}}\right),
\label{eq:total}
\end{equation}
where $e$ is Euler's number.
The frequency of engineering (make) interactions is then calculated as
\begin{equation}
{\rm E}\{p_\rr{m}\} = \frac{\eta}{\mathcal{S}_{\bm P}\left(1 + \eta - \frac{\eta}{e}\right)^2}.
\end{equation}
Finally the frequency of the non-interaction is calculated as $\rr{E_{SS}}\{p_\rr{\varnothing}\} = 1 - \rr{E_{SS}}\{p_\rr{e}\} + \rr{E_{SS}}\{p_\rr{n}\}$ and  $\rr{E_{SM}}\{p_\rr{\varnothing}\} = 1 - \rr{E_{SM}}\{p_\rr{e}\} + \rr{E_{SM}}\{p_\rr{n}\}+ \rr{E_{SM}}\{p_\rr{m}\}$ for species-species and species-modifier interactions, respectively.
Pairwise interactions are assigned randomly from these probabilities between species-species and species-modifiers independently in both quadrants, such that the source pool matrix has no imbued structure apart from the number of species, the number of modifiers, and the frequency of each directional interaction type.
Each source pool is provided a \emph{basal resource} (the first row/column).
A species with a trophic interaction to this resource is identified as an autotroph (or mixotroph depending on its other trophic interactions).
If they do not have service dependencies with other species/modifiers, it is these species that are uniquely able to initiate assembly.

\begin{figure*}[h!]
\centering
\includegraphics[width=0.8\textwidth]{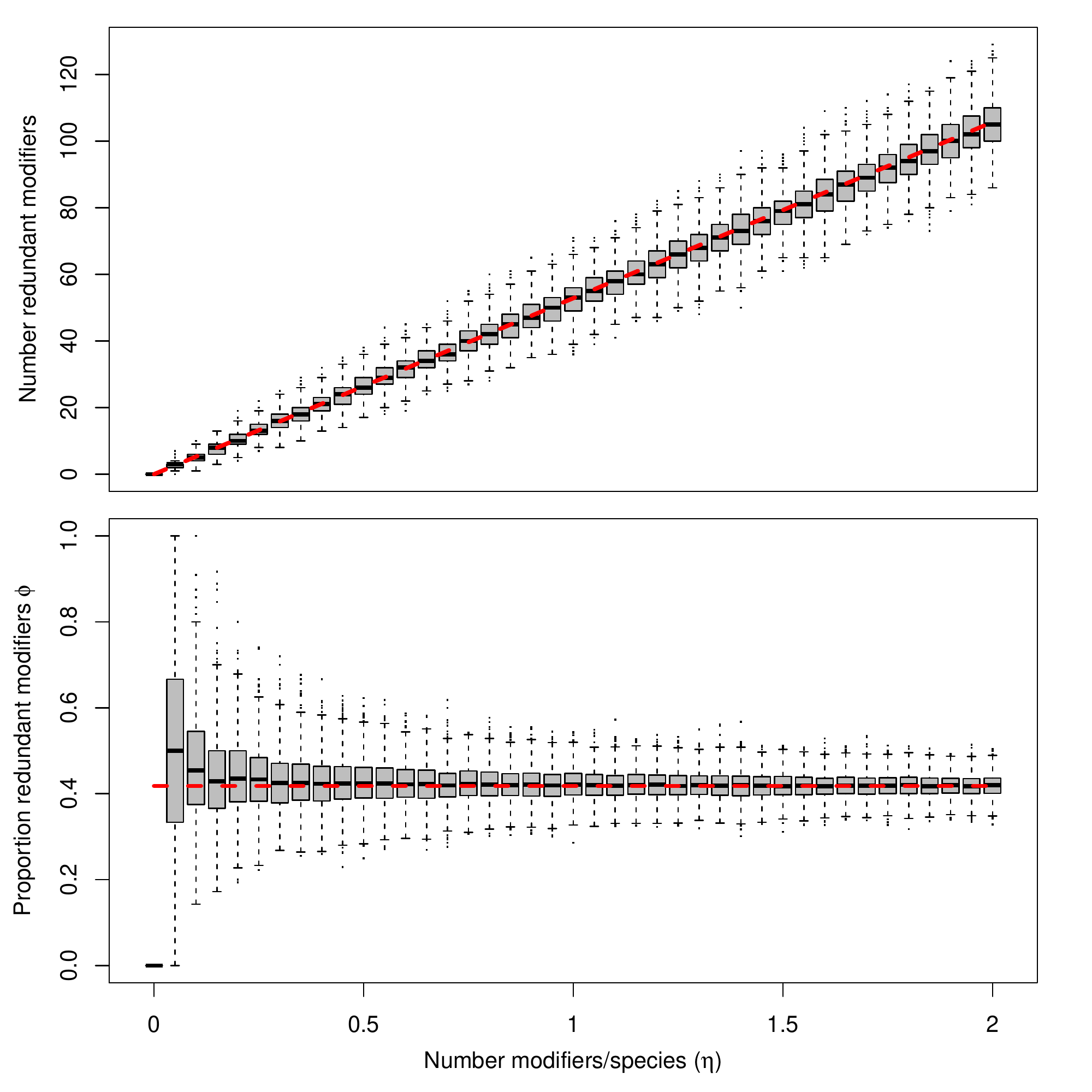}
\caption{
A. Number of redundant modifiers in the source pool as a function of the expected number of modifiers made per species $\eta$.
The red dashed line shows the analytical expectation (Eq. \ref{eq:redundant}).
B. Proportion of redundant modifiers $\phi$ versus the total number of modifiers in the source pool as a function of the expected number of modifiers made per species $\eta$.
The red dashed line shows the analytical expectation of $\phi \approx 0.418$ (Eq. \ref{eq:redundantprop}).
}
\label{fig:redundancy}
\end{figure*}

We can determine analytically the expected number of unique versus redundant modifiers in the source pool.
As the total number of modifiers is given in Eq. \ref{eq:total}, the number of unique modifiers is given by ${\rm E}\{\mathcal{M}_{\bm P}\}_{\rr{unique}} = \mathcal{S}_{\bm P} \eta e^{-1}$.
The number of redundant modifiers is then given as
\begin{equation}
{\rm E}\{\mathcal{M}_{\bm P}\}_{\rr{redundant}} = \eta \mathcal{S}_{\bm P} \frac{e - 2}{e},
\label{eq:redundant}
\end{equation}
such that the proportion of redundant modifiers $\phi$ is
\begin{equation}
\phi = \frac{e-2}{e-1} \approx 0.418.
\label{eq:redundantprop}
\end{equation}
Accordingly, we find that the number of redundant modifiers increases linearly with $\eta$, while the proportion of modifiers that are redundant is fixed.
Figure \ref{fig:redundancy}A,B shows both analytical expectations and numerically-derived measures for ${\rm E}\{\mathcal{M}_{\bm P}\}_{\rr{redundant}}$ and $\phi$, respectively.

\subsection*{Section II: Comparison to niche model}
We compared certain structural features of ENIgMa at steady state to those of the Niche Model \cite{Williams2000}.
Comparisons were restricted to networks constructed in the absence of engineering because engineers introduce indirect effects that are not considered in static food web models, and may make such comparisons irrelevant.
While there are many similarities, there are also some important differences, some of which are highlighted in the main text.
While we consider a comparison of our framework with other food web models such as the Niche Model relevant, we emphasize that the motivations underlying both are distinct.
Our approach is intended to provide a deeper understanding into how multitype dependencies between species and the environment impact the dynamics of community assembly.
While capturing general qualitative features of empirical systems demonstrates that the dynamics we consider are ecologically relevant, the goal of our approach is distinct from that of static food web models, which aim to maximize structural similarities between model and empirical systems \cite{Williams2000,Williams2011}.

We compared steady state ecological networks that emerge from ENIgMa (described in Materials and Methods, main text) with food webs constructed from the Niche Model \cite{Williams2000} with similar species richness and connectance.
Because species richness and connectance of the Niche Model are often altered by eliminating disconnected species, we compared
\emph{i}) species richness,
\emph{ii}) connectance,
\emph{iii}) mean species degree,
\emph{iv}) standard deviation of out-degree distributions, and
\emph{v}) standard deviation of in-degree distributions
averaged across 1000 replicates for each model.

\begin{figure*}[h!]
\centering
\includegraphics[width=0.8\textwidth]{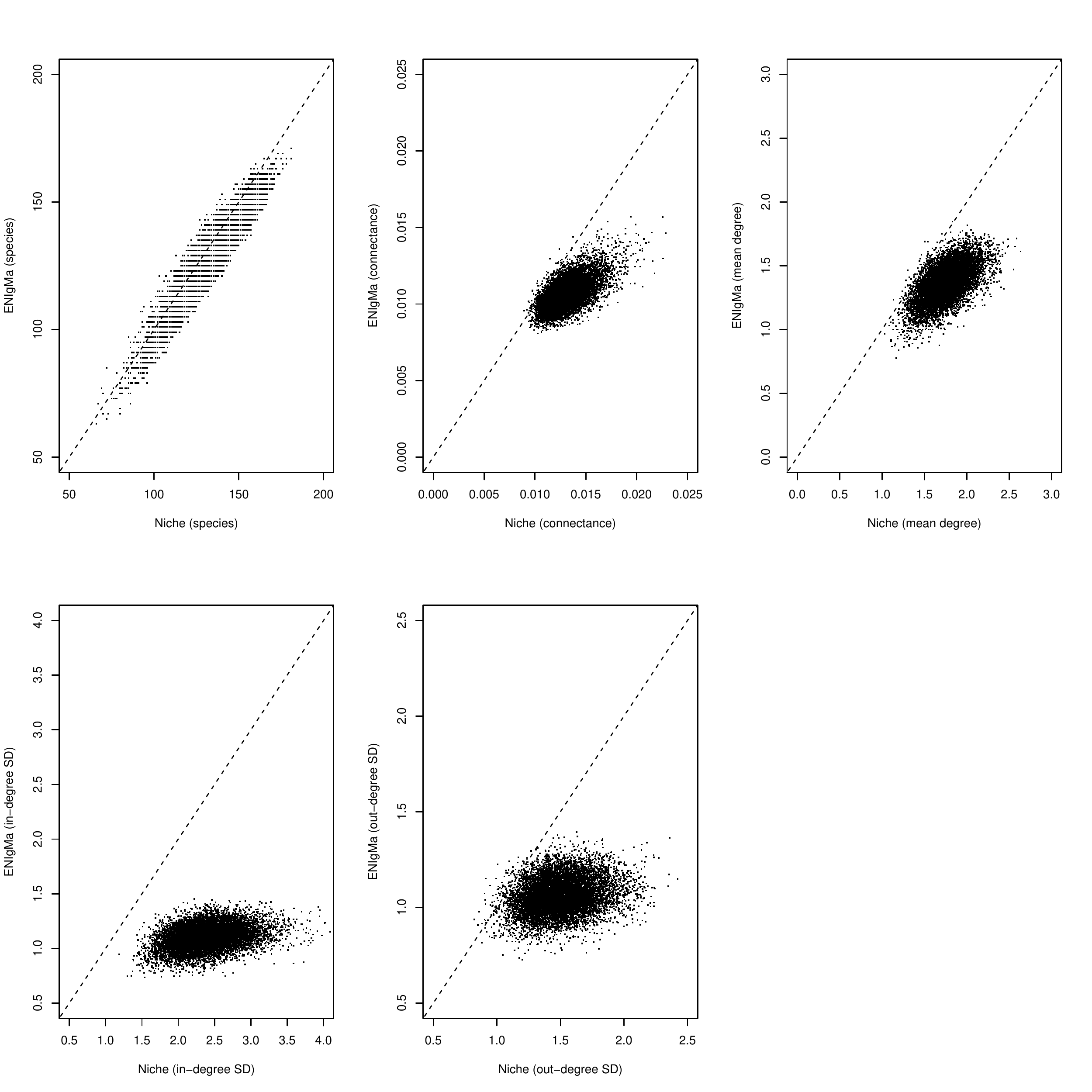}
\caption{
Comparisons of raw structural measures for the assembly (y-axis) and Niche model (x-axis).
If the models produce similar structures, metrics will tend to fall on the 1:1 line (drawn).
While the values for both models are similar, connectance, mean degree, and the standard deviation of in- and out-degree are all lower for the assembly model relative to those measures for the Niche model.
}
\label{fig:error1}
\end{figure*}

\begin{figure*}[h!]
\centering
\includegraphics[width=0.3\textwidth]{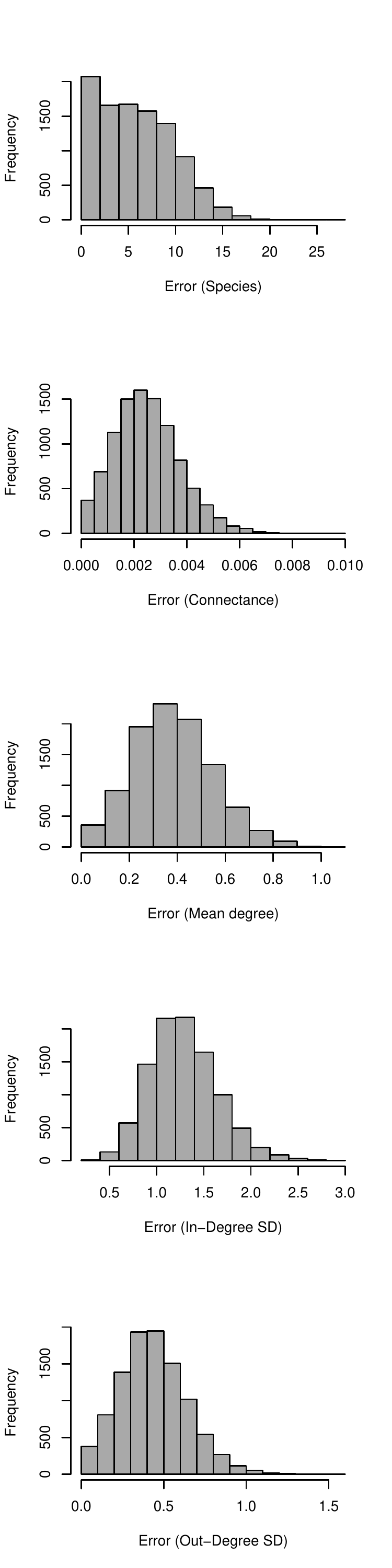}
\caption{
Error between structural measures of the assembly and Niche models.
Error is measured as $\sqrt{(m_i - m_j)^2}$, where $m_i$ and $m_j$ are structural metrics for the assembly and Niche model, respectively.
Only the trophic network of the assembly model used to assess metrics.
}
\label{fig:error2}
\end{figure*}

We found that all measures resulted in fairly similar values between ENIgMa and the Niche Model food webs with a some important differences (Figs. \ref{fig:error1},\ref{fig:error2}).
While similar, ENIgMa produces consistently lower values of connectance, mean species degree, as well as standard deviations of the in- and out-degree distributions.
This means that the food webs produced by ENIgMa are more sparsely connected with less variance between species.
These results were expected, as the Niche Model assumes systematically increasing dietary ranges with higher niche values, whereas the trophic interactions assigned to species in the source pool of ENIgMa are drawn independently.
An important difference between the Niche Model and ENIgMa is that we do not distinguish between predators and parasites.
A different framework known as the Inverse Niche Model \cite{Warren2010} has been proposed to address parasitic interactions.
The Inverse Niche Model assumes increasing specialization with feeding hierarchies, which would serve to lower the average generality of species (lower degree).
In addition, the Inverse Niche model outputs lower standard deviations of in- and out-degree distributions.
Together these trends suggest that the qualitative structural differences that we observe for the assembly and Niche model may reflect an important structural distinction between food webs that do and do not include parasitic species.

\subsection*{Section III: Measures of generality}

The trophic breadth of potential colonizers is thought to play an important role in community assembly.
The definition of a specialist or generalist to some degree depends on the size and connectance of the larger food web.
Trophic generality for a species $i$ is defined $G_i = k^{\rm in}_i/(L/S)$, where $k^{\rm in}_i$ is the in-degree, or number of resources consumed by species $i$ \cite{Williams2000}.
A species is classified as a generalist if the number of its trophic interactions is greater than the average number of links per species, or $G_i > 1$, and a specialist if $G_i < 1$, where a community can be described by the proportion of specialists found therein. 
For interaction networks that are assembling over time, generality can be scaled by a number of different measures of $L/S$, and this has a large effect on our interpretation of the role of generality in community assembly.
For instance, $L/S$ may be quantified by either including all autotrophic species or only autotrophic functional groups.
Furthermore, the scaling of generality may be made with respect to the current state of the community at each point in time, or with respect to the community at steady state.
For instance, in their investigation of assembling mangrove food webs, Piechnik et al. \cite{Piechnik2008} scaled trophic breadth to a standard steady state value of $L^*/S^* = 0.2$ averaged across 102 food webs.

\begin{figure*}[h!]
\centering
\includegraphics[width=0.8\textwidth]{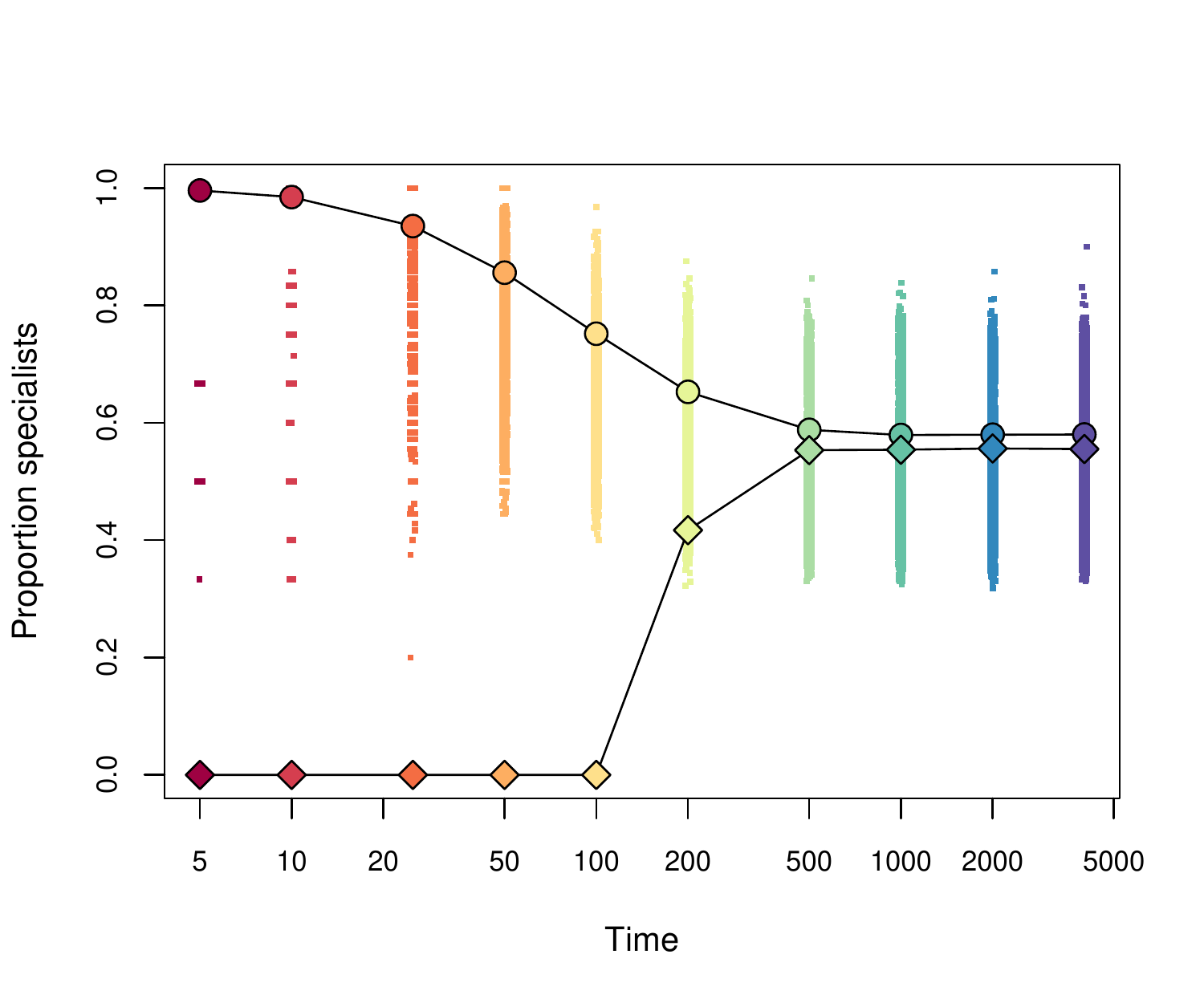}
\caption{
The proportion of specialists as a function of assembly time, where a specialist is defined as a species with a generality index $G_i < 1$.
Measures of $G_i$ are shown normalized to different measures of link-density.
Circles: $G_i^{\rr{all}}$ where $L$ accounts for all links in the food web and $S$ accounts for all species relative to each time interval in the assembly process (averaged across replicates).
Points: $G_i^{\rr{hetero}}$, where we consider only the links and species richness of heterotrophs, excluding autotrophs (each point shows an individual replicate).
Diamonds: $G_i^*$, where $L$ and $S$ are measured with respect to the communities at steady state (averaged across replicates). 
This measure is the one presented in the main text and most similar to that used to evaluate assembling mangrove food webs \cite{Piechnik2008}.
}
\label{fig:spec}
\end{figure*}

To examine how our assessment of the role of generalism over the course of assembly changes based on the application of different scalings, we employ three different measures of $L/S$ to calculate $G_i$:
1) $G_i^{\rr{all}}$, where $L$ accounts for all links in the food web and $S$ accounts for all species relative to each time interval in the assembly process (circles; Fig. \ref{fig:spec}b);
2) $G_i^{\rr{hetero}}$, where we consider only the links and species richness of heterotrophs, excluding autotrophs (points; Fig. \ref{fig:spec}b);
3) $G_i^*$, where $L$ and $S$ are measured with respect to the communities at steady state, which is most similar to the measure used to evaluate assembling mangrove food webs (diamonds; Fig. \ref{fig:spec}b).
Whether trophic breadth is scaled to the current state of $L/S$ or the steady state value of $L^*/S^*$ has a large influence on the estimated proportion of generalists in the community, particularly when the size of the system is small.
We observe that for $G_i^{\rr{all}}$, the system is initially assembled by specialist species, though over the course of assembly the proportion of specialists relative to generalists declines to intermediate values (circles representing the average over replicates in Fig. \ref{fig:spec}).
If only the trophic links between non-autotrophs are considered as in $G_i^{\rr{hetero}}$, specialists still dominate early in assembly, but there is a greater range, such that some systems can be described by a mixed proportion of specialists and generalists (individual points representing independent replicates in Fig. \ref{fig:spec}).

The different normalizations by which generality is measured will impact the interpretation of both empirical and model systems alike.
In our framework, species colonizing early in the assembly process are generalists compared to how the term is defined at the steady state, but they are functionally specialists with respect to the assembling community.
For example, a species that is trophically connected to 10 resource species in the source pool may colonize a community where it is consuming a small subset of its potential range.
As the community grows, that species may realize more of its trophic niche if those resource species subsequently colonize the system.
To what end we label this species a generalist or specialist relative to the assembling community is thus subject to multiple interpretations.

\end{document}